# Topological passivation makes high strength alloys insensitive to hydrogen embrittlement

Huijie Cheng[1], Binhan Sun[1*], Aochen Zhang[1], Dirk Ponge[2], Fengkai Yan[3], Tiwen Lu[1], Xian-Cheng Zhang[1*], Dierk Raabe[2*], Shan-Tung Tu[1]

[1]Key Laboratory of Pressure Systems and Safety, Ministry of Education, East China University of Science and Technology, Shanghai, 200237, China;

[2]Max Planck Institute for Sustainable Materials, Max-Planck-Straße 1, Düsseldorf, 40237, Germany;

[3]Shenyang National Laboratory for Materials Science, Institute of Metal Research, Chinese Academy of Sciences, Shenyang, 110016, China.

*Corresponding authors, E-mails: binhan.sun@ecust.edu.cn, xczhang@ecust.edu.cn, d.raabe@mpie.de.

**Abstract:**

Infrastructure parts for a hydrogen (H) economy need alloys that are mechanically strong and at the same time resistant to the most dangerous and abrupt type of failure mode, namely, H embrittlement. These two properties are in fundamental conflict, as increasing strength typically amplifies susceptibility to H-related failure. Here, we introduce a new approach to make alloys resistant to H embrittlement, by creating a topological passivation layer (up to a few hundred micrometers thick) near the material surface, the region that is most vulnerable to H ingress and attack. The approach is fundamentally different from conventional passivation methods against environmental attack which involve surface chemical reactions. In contrast, topological passivation works without chemical modification or coating. It features instead a layer of ultrafine laminated grains with tens of times higher dislocation density than conventional materials, altering H diffusion, trapping and crack evolution. We tested the concept on a face-centered cubic (FCC) CoCrNi medium entropy model alloy which undergoes severe H-induced intergranular cracking. Two key mechanisms create the topological passivation: First, the high density (up to $\sim 1.3 \times 10^{15}$ m$^{-2}$) of H-trapping dislocations within the passivating grain layer decelerates H migration by up to about an order of magnitude, delaying H-induced crack initiation at grain boundaries. More importantly, once unavoidable micro-sized H-induced intergranular cracks emerge in the topmost surface region, they become completely arrested by the laminated grains, due to a transition in the embrittlement mechanism from H-enhanced grain boundary decohesion to highly energy-dissipative dislocation-associated cracking. These effects almost completely eliminate H embrittlement, at even doubled yield strength, when exposing the so architected material to harsh

H attack. Our approach leverages surface mechanical treatments to tailor metallic microstructures in surface regions most susceptible to H attack, providing a scalable solution to protect alloys from H-induced damage.

**Main Text:**

Sustainably produced hydrogen (H) is an energy carrier that can help mitigating carbon dioxide emissions and reliance on fossil fuels[1]. A crucial step toward advancing a H-fueled economy is the development of lightweight infrastructure components capable of reliably operating in environments with high H fugacity. This is jeopardized by H embrittlement, a phenomenon causing abrupt and complete loss of a material's load-bearing capacity as a result of the interaction between ingressive H atoms and lattice defects such as vacancies, dislocations and interfaces[2, 3]. Due to the slow diffusion of H in alloys with face-centered cubic (FCC) lattice structure and their high toughness, FCC alloys like austenitic stainless steels are usually less affected by H embrittlement compared to body-centered cubic (BCC) alloys, e.g., carbon steels[4], rendering them promising candidates for components used in harsh H containing environments[5]. However, precipitate-free FCC materials (austenitic stainless steels, Ni-based materials and medium/high entropy alloys) often show a relatively low yield strength (typically below 400 MPa[6, 7]), a factor which limits their use for strong and lightweight components. Thus a fundamental challenge exists to reconcile high strength and excellent H resistance, two so far mutually excluding essential features, in one material system[8].

Mechanically stable FCC alloys against deformation-induced phase transformations often fracture in an intergranular mode in the presence of H. This is closely linked to the segregation of H to high-angle grain boundaries before or upon loading and the subsequent reduction in interface cohesive strength[9]. This effect is called H-enhanced decohesion (HEDE)[10]. Microstructure strategies designed to suppress (but not eliminate) H-induced intergranular cracking in these materials, thereby enhancing their H embrittlement resistance, typically focus on reducing H accumulation at grain boundaries. This can be accomplished by introducing H-trapping precipitates to suppress internal H migration[11], or by grain refinement to increase the grain boundary area per unit volume (i.e. the reduction of the grain boundary H coverage)[12]. However, these measures do not ultimately prevent intergranular cracking and lose efficacy in components subjected to continuous H exposure, as H saturation at grain boundaries eventually occurs[13]. Therefore, a more effective method that can fundamentally alter both, H migration and trapping on the one hand and at the same time H-induced damage percolation on the other, is needed to enhance H tolerance of FCC materials.

Here, we present a novel microstructure-based approach that directly targets the most H-susceptible region of materials – the surface – by generating a topological passivation layer. This layer functions to (a) suppress H migration and accumulation and (b) mitigate H-induced cracking initiated at or near the surface. To demonstrate our concept, we utilize an equimolar CoCrNi medium-entropy alloy (MEA) as a mechanically stable model face-centered cubic (FCC) material.

Unlike traditional passivation based on surface chemical reactions to form oxides against corrosion attack[14], topological passivation does not require any changes in chemistry. Instead, it is characterized by a near-surface layer of a few hundred micrometers thickness of ultrafine laminated grains containing a high density of dislocations (up to ~$1.3×10^{15}$ $m^{-2}$, Fig. 1a). The dislocations effectively trap the incoming H and thus decelerate H ingress, migration and accumulation at the vulnerable grain boundaries. Moreover, the H-induced intergranular cracks, nucleated at the topmost surface region, are completely arrested by the pancake-like topological alignment of the laminated grains. This crack arresting effect is governed by a transition in the underlying H embrittlement mechanisms, shifting from a sudden brittle decohesion, as described by HEDE, to a highly energy dissipating process involving massive dislocation activity. The latter is characterized by a sequence of dislocation emission, crack blunting, dislocation-interface interactions, and the nucleation and coalescence of new sub-critical cracks. These effects, associated with the pancake-like topological features of the passivation layer, provide near-complete resistance to H embrittlement at nearly doubled yield strength when exposing the architected material to harsh electro-chemical H charging conditions.

The microstructure-based topological passivation approach offers several fundamental advantages, including scalability, low energy requirements, and the elimination of chemical surface treatments. Other bulk processing methods such as thermomechanical treatments to produce fine grained microstructures against H embrittlement are costly, due to the heat involved. Also, they can leverage unwanted effects such as reduced formability or they suffer from component size limitations. In contrast, our approach creates topological passivation layers through shearing the sample surface via a severe mechanical treatment using an ultrasonic surface rolling process (USRP), shown in Fig. 1a.

The layer resulting from this process has a thickness of ~150 μm in the architected CoCrNi MEA (Figs. 1b and c). The topmost ~10 μm of this compound layer region consists of nano-sized equiaxed grains with an average size of ~100 nm, as shown in Fig. 1d. The applied ultrahigh shear strain (up to ~56.8) and strain rate (up to ~$5×10^{3}$ $s^{-1}$) imposed on the sample surface by USPR promote the generation of dislocations and suppress their annihilation by recovery[15], leading to the occurrence of grain recrystallization that dissipates the stored energy of deformation[16, 17]. In the subsurface region spanning from ~10 to ~150 μm, constituting the majority of the passivation layer, the shear strain and strain rate acting on the material are lower[18]. Consequently, the fraction of recrystallized grains decreases, and the non-recrystallized grains become elongated as a response to the shear (Fig. 1e). The average lamellar thickness and aspect ratio of these laminated grains in the passivation layer are 2 μm and ~4.2 (see the inset in Fig. 1e), respectively. Electron backscatter diffraction (EBSD) and electron channeling contrast imaging (ECCI) analysis (Figs. 1e and f) reveal the presence of high densities of dislocations, deformation twins as well as low-angle grain boundaries (LAGBs) that are formed due to dislocation recovery within the laminated grains[19]. Beyond a depth of approximately 150 μm (i.e., outside the passivation layer), the USRP-induced shear strain and strain rate are no longer sufficient to significantly refine or heavily elongate the coarse grains. Only some deformation substructures, such as deformation twins and dislocation slip bands, are formed in the grain interior close to the passivation layer (Supplementary Fig. 1). Fig. 1g shows the distribution of nanohardness at varying depth counted from the sample surface. The hardness increases from initially ~4.7 GPa at the surface region to ~5.0 GPa at a depth of ~100 μm (within the subsurface laminated grain layer), followed by a continuous decrease towards regions away from the passivation layer. The initial increase in nanohardness is attributed to the buildup in dislocation density in the laminated grain layer (Fig. 1g), resulting from incomplete recrystallization. The distribution of the compressive residual stress follows the same trend as the hardness (Fig. 1g). It gradually increases from ~10 MPa at the sample interior (300 μm depth from the surface) to a maximum value of 381 MPa at a depth of 80 μm (subsurface laminated grain layer) due to the increase in accumulated plastic strain, followed by a decrease to 312 MPa at the top surface owing to the surface relaxation[20].

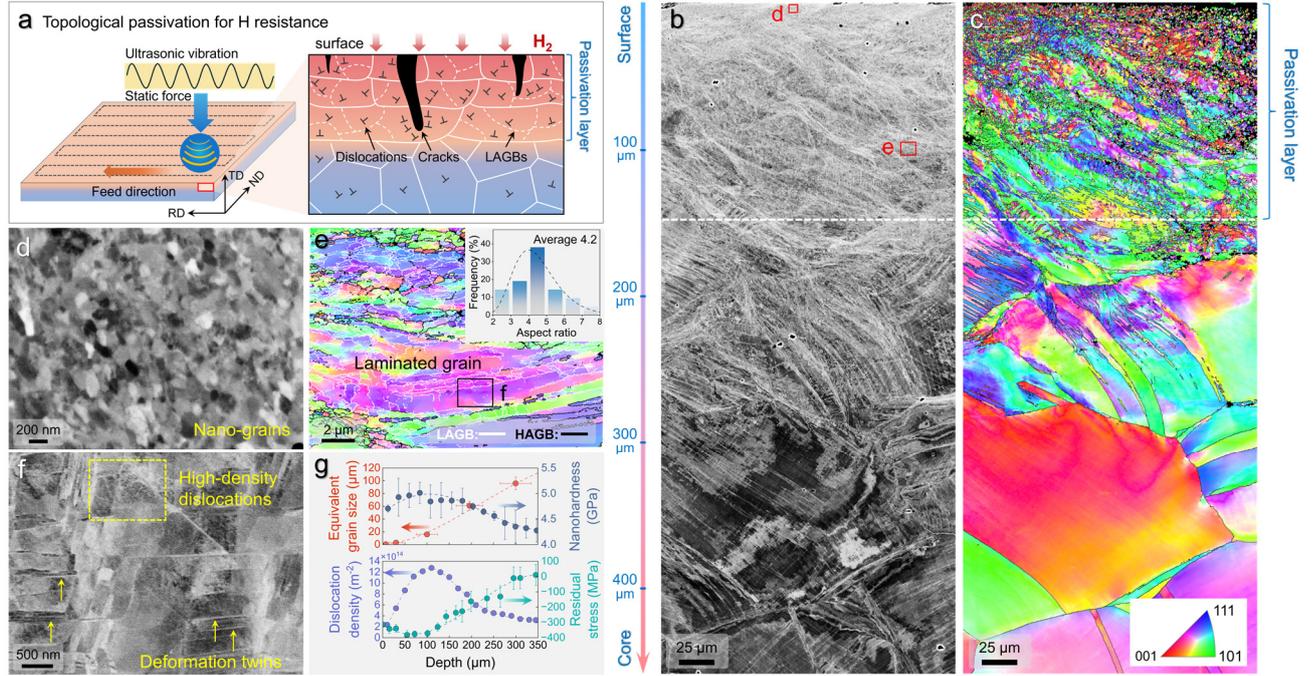

**Fig. 1 Microstructure of the H-resistant CoCrNi alloy after topological passivation.** (a) Schematic illustration of our topological passivation concept, which uses surface mechanical treatment to form a microstructure layer containing highly deformed and laminated grains to suppress H-induced damage. (b) Electron channeling contrast (ECC) image showing the longitudinal-sectional overview of the microstructure spanning from the mechanically treated surface to a depth up to ~450 μm. (c) Electron backscatter diffraction (EBSD) inverse pole figure (IPF) map corresponding to the same area as shown in b. (d) ECC image taken from the surface region (corresponding to a depth of ~10 μm) as marked in b, showing the equiaxed nano-sized grains. (e) EBSD-IPF map taken from the subsurface layer roughly marked in b (corresponding to a depth of ~100 μm), showing the representative laminated grained microstructure. The inset is the distribution of the aspect ratio of the laminated grains in the subsurface layer (estimated within a depth span of 10-150 μm). (f) High-magnification ECC image showing detailed deformation microstructure within the laminated grain marked in e. (g) Distribution of nanohardness, equivalent grain size (measured from EBSD), dislocation density (calculated based on nanohardness data, Supplementary Note 1), and residual stress (measured by X-ray diffraction) at different depths from the sample surface. LAGBs/HAGBs: low/high-angle grain boundaries.

Our approach of introducing a topological passivation layer results in a notably improved resistance to H embrittlement. This is demonstrated by comparing the passivated sample to a reference specimen that has a homogeneous coarse-grained (CG, average grain size ~140 μm) microstructure (without USRP processing, hereafter referred to as CG sample). Both samples were subjected to the same *in-situ* H charging conditions (see Materials and Methods) during the slow strain-rate ($5 \times 10^{-6}$ s$^{-1}$) tensile testing. We used an electrochemical charging condition with a H fugacity equivalent to that of ~100 bar gaseous H$_2$[21], which exceeds the pressure conditions of most pipelines and tubes for H transport[22]. Under such a severe environment, the topologically passivated sample exhibits near-complete resistance to H embrittlement, with only a modest H-induced reduction in tensile elongation (Fig. 2a and Supplementary Fig. 2, 0.2%~3.5%). This H-induced ductility loss is much smaller than that of the CG sample, which experiences a substantial ductility loss of 15.5%~19.6%. Moreover, the passivation layer increases the overall yield strength of the sample by ~360 MPa compared to the CG sample (yield strength ~340 MPa). That is, our strategy can simultaneously increase strength and H resistance, a synergistic result exceeding the results from other strengthening approaches (e.g., cold deformation, Fig. 2b) that often result in a trade-off between strength and H resistance.

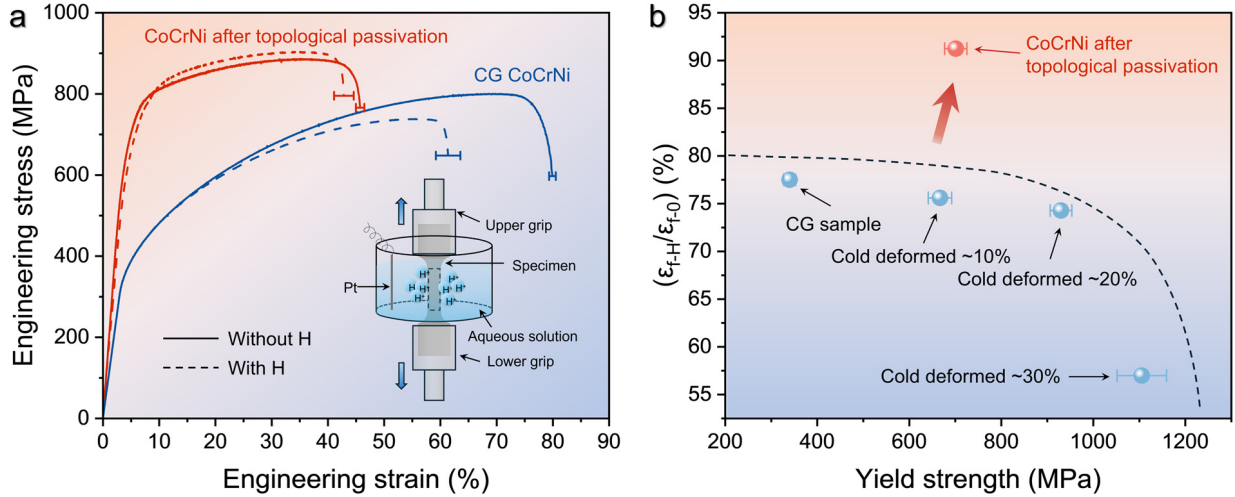

**Fig. 2 Improvement in the resistance to H embrittlement via topological passivation.** (a) Slow strain-rate tensile properties of the topologically passivated CoCrNi sample in comparison to the CG reference material under *in-situ* H charging. The thickness of both samples is 1.2 mm. The inset shows the schematic of the *in-situ* electrochemical H-charging setup (current density 20 mA cm$^{-2}$, room temperature). (b) The H embrittlement index (the ratio of ductility between the samples tested with H ($\varepsilon_{f\text{-}H}$) and without H ($\varepsilon_{f\text{-}0}$), i.e., $\varepsilon_{f\text{-}H}/\varepsilon_{f\text{-}0}$) as a function of yield strength for the topologically passivated sample, in comparison to reference materials that are strengthened by different levels of cold deformation. CG: coarse-grained.

To better understand the underlying mechanisms of the improved H embrittlement resistance in the topologically passivated CoCrNi alloy, we begin by comparing the fractography of both samples tested and shown in Fig. 2a. In the absence of H, both samples show a complete dimple-type ductile fracture behavior (Supplementary Fig. 3). The CG CoCrNi sample is clearly embrittled when exposed to H, with nearly 41% of the fracture surface showing distinct intergranular fracture behavior (Figs. 3a and b). This is attributed to H-induced cracking along austenite grain boundaries, as shown from the EBSD analysis in Fig. 3c. This cracking mode, also observed in other stable FCC alloys such as austenitic stainless steels[23], high-Mn steels[24], pure Ni and Ni-based superalloys[25, 26] and other medium/high entropy alloys[27, 28], can be explained through the HEDE mechanism. The segregation or respectively trapping of H atoms at austenite grain boundaries ultimately reduces their cohesive strength to a level lower than the critical stress required for dislocation emission (a crack blunting mechanism), a circumstance which favors crack nucleation and propagation[29, 30]. This process can be further enhanced by stress concentrations at grain boundary regions that are naturally developed due to elastic and inelastic misfit between adjacent grains and the associated formation of geometrically necessary dislocations during deformation (Supplementary Fig. 4).

In contrast to the reference CG sample, the topologically passivated sample exhibits a significantly smaller H-affected zone, covering only 6.5% of the fracture surface area (Fig. 3d). In fracture regions closest to the specimen surface, a pronounced 'rugged' fracture morphology is observed (Figs. 3e and h). A more detailed scanning electron microscopy (SEM) observation reveals the presence of fine intergranular facets (average size 100 nm) within this region (Fig. 3g). This observation indicates the occurrence of H-induced grain boundary cracking along the equiaxed nanograins at the topmost surface. This is further validated by a cross-sectional characterization of the near-surface fracture region (realized by a combined focused ion beam (FIB) and transmission electron microscopy (TEM) technique), where H-induced secondary cracking along grain boundaries is observed (Figs. 3i and j). This finding suggests that even though the grain size has been refined to tens or hundreds of nanometers, the predominant H-induced damage mode remains unchanged. However, in deeper regions within the passivation layer containing laminated grains, the H-induced intergranular fracture changes to a quasi-cleavage fracture manner featuring large facets with some ridges (Fig. 3f). This distinct transition in the fracture mode implies a change in H-induced cracking mechanisms, which will be discussed in more detail below. Next, we investigate the H-induced cracking behavior of the passivated CoCrNi alloy by characterizing the fractured specimens from a longitudinal section perspective (Fig. 4a). We observe that most H-induced microcracks, which intruded from the sample surface regions where H ingress occurs most rapidly, get strongly blunted and successfully arrested within the laminated grained layer (depth below ~45 μm, Supplementary Fig. 5). One representative crack, as demonstrated in Figs. 4b and c, clearly shows such an arresting effect that laminated grains exert on H-induced crack propagation, establishing a key factor contributing to the improved H resistance of the topologically passivated material.

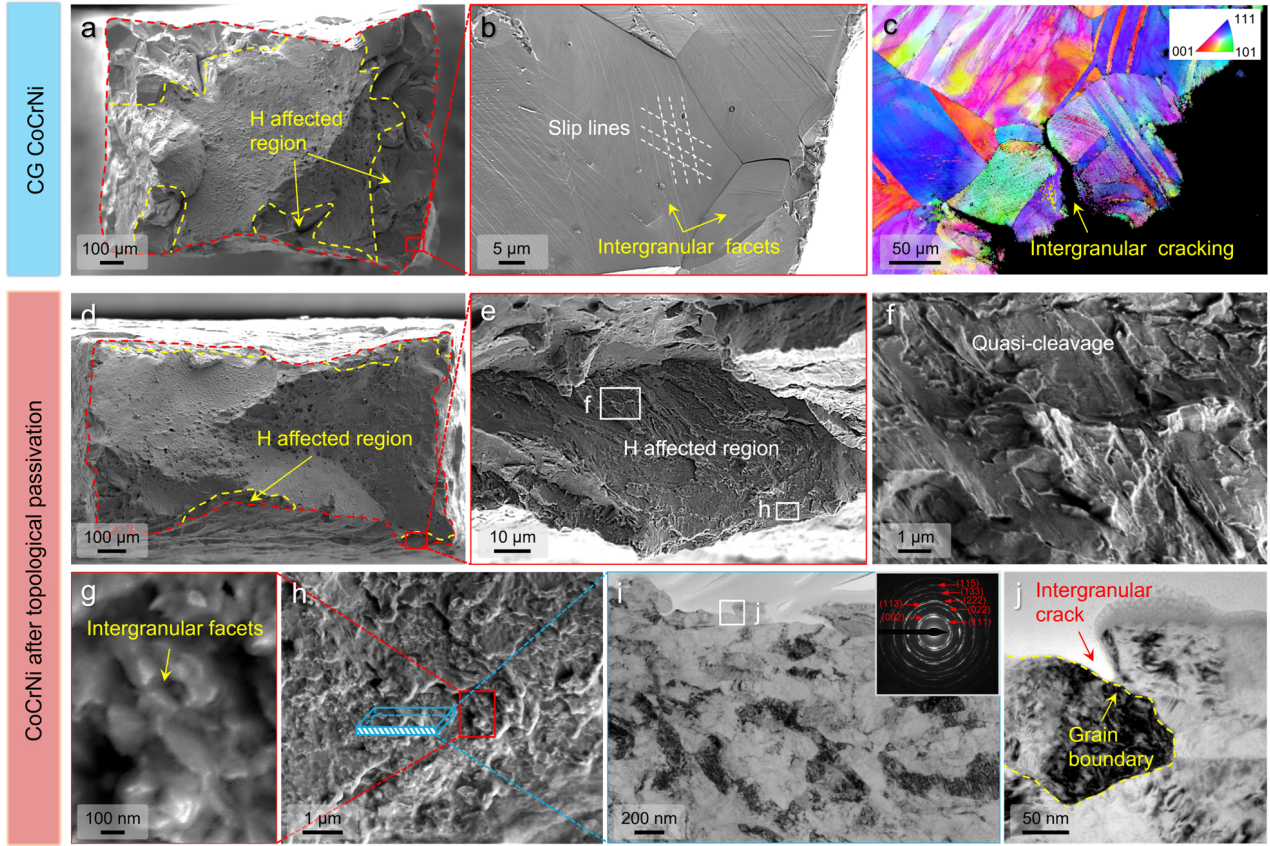

**Fig. 3 H-induced fracture behavior of the investigated materials.** (a) Fracture surface of the reference CG sample tested under H, showing clear H embrittlement. (b) Magnified SEM image taken from the H affected region marked in a, showing intergranular fracture features. (c) Representative EBSD-IPF image of a region near the fracture surface, showing H-induced cracking along the grain boundaries. (d) Fracture surface of the topologically passivated sample tested under H environment, showing a much smaller H-affected zone compared to the CG sample. (e) Magnified SEM image taken from the rectangular frame marked in d, showing a representative fracture morphology within the H affected region. (f) SEM characterization from the subsurface region (at a depth of ~40 μm) marked in e, showing quasi-cleavage fracture behavior. (g)-(h) Magnified SEM images taken from the surface region (at a depth of ~3 μm) marked in e, showing intergranular fracture in the near-surface nanograined layer. (i) Cross-sectional transmission electron microscopy (TEM) bright field (BF) micrograph taken from the subsurface region of the rectangular frame marked in h. The inset shows the selected area electron diffraction (SAED), validating the presence of nanosized grains. The TEM specimen was extracted by the focused ion beam technique performed on the blue-frame region marked in h. (j) Magnified image taken from the rectangular frame marked in i, showing the secondary H-induced crack along the nanosized austenite grain boundaries. CG: coarse-grained.

The activation of HEDE and the resulting nucleation of H-induced intergranular cracking require a critical H concentration to accumulate at grain boundaries[26, 31]. The H diffusion and accumulation process is kinetically influenced by the density of dislocations and grain boundaries[32, 33], two most widely reported H trapping sites that are more abundant in the topological passivation layer (Fig. 1g). Based on the depth-dependent dislocation density and grain boundary fraction (Supplementary Fig. 6, estimated from the nanohardness results and EBSD data, respectively) in the topologically passivated sample, we calculate the total H concentration of this sample as a function of charging time by incorporating their respective H trapping effects[34] (Supplementary Fig. 7, with calculation details provided in Supplementary Note 2). Through fitting such calculation results and the experimental data acquired by thermal desorption spectroscopy (TDS), the H diffusivity at various depths of the topologically passivated sample can be resolved (Supplementary Fig. 8). We find that the presence of the high defect density within the passivation layer slows the H diffusivity in these regions down to ~0.09 times that of the reference CG sample. Note that this value is the upper bound of the retardation effect for H motion, given that the calculation assumes the upper bound of the defect trapping energy reported in literature[35, 36] and did not consider the documented yet debatable grain boundary diffusion of H[37, 38]. This retardation effect on H transport will correspondingly reduce the accumulation rate of H at grain boundaries and thus the initiation of H-induced cracks.

However, once H-induced cracks are nucleated at the surface region, the fresh cracking surface will be immediately exposed to the applied H atmosphere which triggers H ingress, particularly into the stress-concentrated crack-tip regions. This suggests that further propagation of these intergranular cracks should not be limited by the H accumulation rate, which is then again primarily governed by the surface H fugacity boundary condition rather than by the one that is internally reduced by the retarded H transport. As such, the suppressed H-induced crack propagation, observed in the subsurface laminated grained layer, must derive from a microstructure effect. First, we checked whether the residual stress induced by the surface treatment (Fig. 1g) increases the H resistance of the material, by reducing the stress intensity factor, thus increasing the threshold for crack propagation[39, 40]. To do this, we performed a stress relief annealing treatment (400 °C, 5h) for the passivated sample to reduce the residual compressive stress at the (sub)surface regions, from an average value of -320 MPa to -65 MPa after treatment (Supplementary Fig. 9), as quantified by X-ray diffraction. When subjected to loading in the same H atmosphere, the mechanical response of the stress-relieved sample does not differ significantly from that of the as-processed passivated material (Supplementary Fig. 9). This result indicates that the residual compressive stress is not the primary factor contributing to the improved H embrittlement resistance of the passivated CoCrNi alloy.

We therefore next focus on the deformation and damage behavior within the laminated grain regions and discuss how the cracks are arrested. Detailed characterization of representative H-induced cracks reveals that the crack propagates through a transgranular mode within the laminated grain regions (as shown in Figs. 4d-f). The transition in cracking mode can be explained in terms of energy dissipation. If crack propagation remains in an intergranular mode, it will undergo a crack deflection of 90° in the laminated grains. Based on linear elastic fracture mechanics analysis[41], this cracking process will result in a more significant reduction of the effective stress intensity factor or cracking driving force (by 50%) compared to a 60° crack deflection for the equiaxed grained structure (only 25% reduction) (Supplementary Note 3). As a result, H-induced transgranular cracking is more energetically favorable within the laminated grains. It is further observed that in front of the crack tip, nanosized globular grains surrounded by high-angle grain boundaries (HAGBs) are formed, as revealed by the transmission Kikuchi diffraction (TKD) results in Fig. 4f. Some of the HAGBs undergo decohesion, triggering the nucleation of new nano-cracks (with a length of only a few nanometers, Figs. 4e-g), which will eventually coalesce with the main crack upon further loading. This observation suggests that the nature of H-induced transgranular cracking within the laminated grains is a process of nanograin formation and its subsequent H-induced decohesion followed by further crack coalescence. Given that these globular nanograins are rarely observed within the laminated grains in the initial microstructure (Fig. 1e), we propose that their formation occurs during deformation under H atmosphere. More specifically, the high density of dislocations formed at the main crack tip (Supplementary Fig. 10), which might also be promoted by the presence of H due to a H-enhanced localized plasticity (HELP)[42] and/or an adsorption-induced dislocation emission (AIDE) mechanism[43], can gradually accumulate at the initially existing low-angle grain boundaries, resulting in strain-induced rotation of subgrains and further development of high misorientations[44, 45]. Such dislocation formation and accommodation process essentially dissipate a significant amount of plastic energy, reducing the applied driving force for crack propagation and thus rendering further crack propagation more difficult[46]. The crack-bridging effect caused by the formation of intergranular nanocracks in front of the main crack will further reduce the crack-tip stress intensity and continuously suppress crack propagation[47]. This crack arresting mechanism is schematically illustrated in Fig. 4j. In addition to this point, we further observe the formation of dense arrays of deformation twins within the laminated grains close to the crack tip (Figs. 4h and i), which might also be promoted by the presence of H due to its effect on reducing the material's stacking fault energy[48]. The coherent twin boundaries can further block the crack advance due to their higher surface separation energies[49] and lower H solubility[50], serving as another reason for the improved resistance to the H-induced cracking propagation.

Unlike bulk processing, which alters the microstructure throughout the entire material, our approach aims to tailor only the microstructure at the surface regions, from where the H enters, turning it into a protective zone. The topological passivation layer is produced by simple surface shear deformation, achievable through a variety of surface treatment techniques (e.g., mechanical attrition[15] and grinding[51]) across different types of materials including pure Ni[18], Cu[15], Al[51] and FeCrAl[52] alloys. To assess the generality of our topological H-resistance strategy, we also tested surface passivated pure Ni, fabricated by using the same shear treatment method as for the CoCrNi alloy. Like for the CoCrNi alloy, complete resistance to H embrittlement is also achieved for pure Ni when exposed to the same harsh *in-situ* H charging condition (Supplementary Fig. 11), in contrast to a substantial H-induced ductility loss (by 70%) observed in Ni reference samples that had not been treated by topological surface passivation.

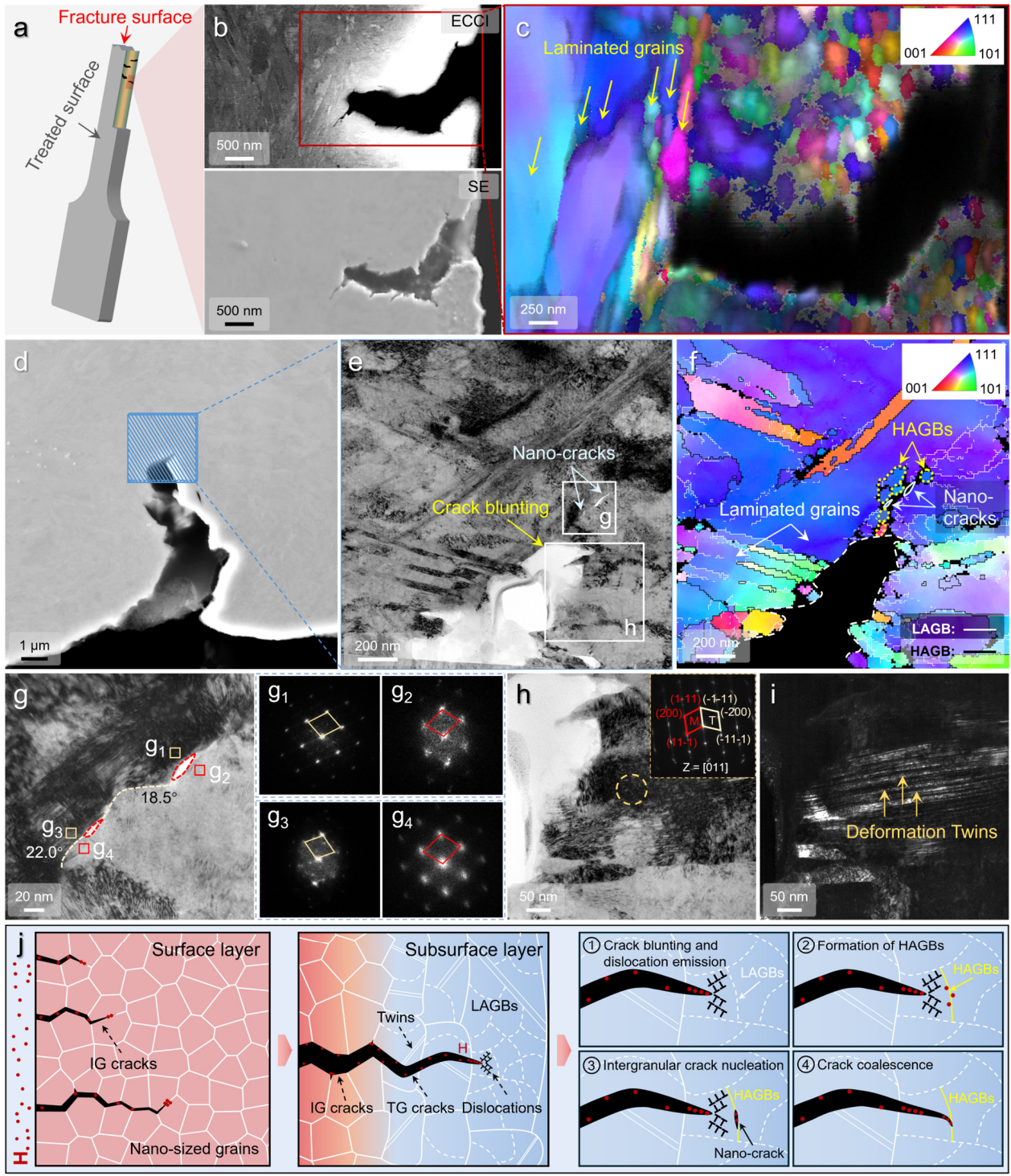

**Fig. 4 H-induced crack arresting mechanisms in the topologically passivated CoCrNi sample.** (a) Schematic diagram of selected regions for characterization in the H-charged and fractured topologically passivated sample.

(b) A representative blunted and arrested H-induced crack probed by secondary electron (SE) and ECC imaging in the H-charged and fractured passivated sample. (c) Detailed EBSD characterization taken from the rectangular frame marked in b, showing a prominent arresting effect of the laminated grains on H-induced crack propagation. (d) A selected blunted and arrested H-induced crack in the H-charged and fractured passivated sample. The crack was probed by SE imaging from the area close to the fracture surface and the specimen's side edge, where H is nearly saturated. (e) TEM-BF micrographs showing the microstructure close to the crack tip. The TEM sample was extracted by FIB performed on the blue-frame region marked in d. (f) Corresponding transmission Kikuchi diffraction (TKD) IPF map of e. (g) Magnified image taken from the rectangular frames marked in e, showing the micro-cracks in front of the crack tip. The misorientation angles of the interfaces that underwent decohesion, determined from fast Fourier transform (FFT) diffraction patterns, is also added. ($g_1$)-($g_4$) Corresponding FFT diffraction patterns taken from the locations marked in g. (h) Magnified TEM-BF image taken from the rectangular frame marked in e, showing the microstructure of deformation twins. The inset shows the selected area electron diffraction, validating the presence of deformation twins. (i) TEM dark field (DF) image corresponding to the same area as h. (j) Schematic illustration showing the initiation, arresting and propagation mechanisms of H-induced cracking within the topologically passivated layer.

In conclusion, we show how to render metallic FCC alloys H-resistant based on architecting their near-surface microstructure, regions otherwise most susceptible to H-induced damage. In addition to suppressing H diffusion, the ultrafine grained layer, particularly when turned into a laminated morphology by surface shear deformation, can effectively prevent H-induced intergranular cracking, the key failure mechanism in FCC metals. Instead, the H-induced cracks are forced to propagate in a transgranular mode via a highly energy-dissipative mechanism based on massive dislocation production, dislocation-grain boundary interactions, new crack nucleation and coalescence processes. These mechanisms, involving intense dislocation motion, cross-slip and multiplication, essentially blunt and successfully arrest H-induced cracks within the passivation layer. This is an essential effect because in metals exposed to permanent H charging, micro-crack formation is practically unavoidable. This approach renders the CoCrNi model alloy (and potentially also other FCC materials) practically immune to H embrittlement under harsh electrochemical H charging conditions. It omits the use of any costly chemical modifications or protective coatings. The topological passivation layer can also be tuned to achieve various combinations of materials' overall strength[53, 54] and be scaled up for application to large, in-service structural components[55, 56], thus offering a versatile and promising solution for the development of safe materials and components used across all sectors that use, store, transport or combust $H_2$.

**Methods**

Alloy fabrication

The CG CoCrNi MEA was produced from high-purity elements (>99.9% pure), which were arc-melted under argon atmosphere. To homogenize the as-cast alloy, the ingot was subsequently subjected to heat treatment at a temperature of 1200 °C for 24 h. Cold rolling with a total thickness reduction of 70% was then performed, followed by an annealing at a temperature of 900 °C for 3 h and water quenching. For the topologically passivated sample, the ultrasonic surface rolling process (USRP) was applied at ambient temperature on the whole gauge section of the tensile specimen. During USRP, a static extrusion force with ultrasonic vibration was imposed on the sample surface. The detailed parameters of the USRP process are: static force 400 N, vibration amplitude 14 μm, rolling speed 2000 mm/min, frequency of ultrasonic mechanical vibration 20 kHz and processing times 25.

Mechanical testing and hydrogen charging

The susceptibility to H embrittlement of the CG and topologically passivated samples was evaluated by *in-situ* H-charging and tensile tests conducted with an initial strain rate of $5 \times 10^{-6}$ s$^{-1}$. Tensile experiments were conducted in an Instron tensile machine. Tensile specimens with a gauge length of 8.5 mm and thickness of ~1.2 mm was used. During the entire duration of tensile testing, *in-situ* electrochemical H charging was performed on tensile specimens in an aqueous solution containing 3 wt.% NaCl and 0.3 wt.% NH$_4$SCN at room temperature. A platinum foil was used as the counter electrode. The total H concentration was controlled by a fixed current density of 20 mA cm$^{-2}$.

Thermal desorption spectroscopy

Thermal desorption spectroscopy experiments were performed using a Bruker G4 PHOENIX to measure the H concentration in pre-charged specimens. Specimens with a dimension of $10 \times 10 \times 1.25$ mm$^3$ were used and were subjected to the same H charging condition as that for tensile specimens. The tests started within 5 min after H charging. The total H concentration was determined by measuring the cumulative desorbed H from room temperature to 800 °C with a constant heating rate of 16 °C min$^{-1}$.

Microstructure characterization

Secondary electron (SE) imaging, electron channeling contrast imaging (ECCI) and electron backscatter diffraction (EBSD) were conducted using a Zeiss Crossbeam 340 and Zeiss Crossbeam 460 scanning electron microscopy (SEM) instrument. Samples for EBSD and ECCI were finally polished with a 0.02 μm colloidal silica suspension. The acquired EBSD data (grain size and local misorientation) were analyzed using the TSL OIM software package. The relationship between area per unit volume of grain boundary ($S_v$) and grain size ($d$) is expressed as $S_v=7.1025\ d^2/d^3$, reported in the work of Mendelson[57]. TEM experiments and selected area electron diffraction (SAED) were carried out in a transmission electron microscopy (TEM, FEI Talos F200X) operated at 200 kV. Transmission Kikuchi diffraction (TKD) analysis was conducted using a Verios 5 UC operated at 20 kV. Thin foils for TEM and TKD were prepared by a site-specific lift-out procedure using a dual-beam focused ion beam instrument (FEI Helios 5 UX).

Nano-indentation tests were performed by a KLA Nano Indenter G200X, using a Berkovich shaped indenter with a load of 50 mN. Residual stress assessment was carried out using a Proto-I X-ray Diffraction (XRD) MG40P FS STD residual stress tester with Mn-Ka radiation at a wavelength of 2.10314 Å. Electrochemical polishing was conducted using the Proto Electropolisher Model 8818 to remove the material layer by layer. The XRD measurement parameters were set to an acceleration voltage of 30 kV and a current of 25 mA. Data analysis was carried out using the sin2φ method[58], with a Bragg angle of 159° and the {311} diffraction crystal plane of the γ-phase.

**Acknowledgments:**
Part of the SEM and EBSD experiments were performed in Liaoning Academy of Materials and we thank Dr. Fanghai Xin for the assistance in performing these experiments. This work was supported by the New Cornerstone Science Foundation through the XPLORER PRIZE, the National Key Research and Development Project (No. 2023YFB3712103, 2022YFB4600019), National Natural Science Foundation of China (No. 52275147), Key Research and Development Program of Shandong Province (No. 2024CXPT064) and Shanghai Gaofeng Project for University Academic Program Development.

**Contributions:** B.S., X.-C.Z. and D.R. conceived the project. H.C. and B.S. designed the experimental/modelling program. H. C. fabricated the material and characterized the microstructure. H. C. and A.Z. performed tensile testing. H.C. and B.S. analyzed the SEM and TEM results. H.C. performed the TDS experiments and H diffusion simulation. B.S., X.-C.Z., D.R. and S.-T.T. supervised the study. H.C. and B.S. wrote the manuscript. D.R. and S.-T.T. revised the manuscript. All the authors discussed the results and commented on the manuscript.

**Competing interests:** The authors declare that they have no competing interests.

**Data availability:** All data are available in the manuscript or the supplementary materials.

**Supplementary information**
Supplementary Note 1-3
Supplementary Figs. 1 to 11
Supplementary Table 1